# FAIR 2.0: Extending the FAIR Guiding Principles to Address Semantic Interoperability

Vogt, Lars[1]  orcid.org/0000-0002-8280-0487;

Strömert, Philip[1]  orcid.org/0000-0002-1595-3213

Matentzoglu, Nicolas[2]  orcid.org/0000-0002-7356-1779

Karam, Naouel[3]  orcid.org/0000-0002-6762-6417

Konrad, Marcel[1]  orcid.org/0000-0002-2452-3143

Prinz, Manuel[1]  orcid.org/0000-0003-2151-4556

Baum, Roman[4]  orcid.org/0000-0001-5246-9351

[1] *TIB Leibniz Information Centre for Science and Technology, Welfengarten 1B, 30167 Hanover, Germany*

[2] *NICO: Semanticly, Athens, Greece*

[3] *Institute for Applied Informatics (InfAI), University of Leipzig, Germany*

[4] *ZB MED - Information Centre for Life Sciences, Gleueler Straße 60, 50931 Cologne, Germany*

Corresponding Author: lars.m.vogt@googlemail.com

—

## Abstract

FAIR data presupposes their successful communication between machines and humans while preserving their meaning and reference, requiring all parties involved to share the same background knowledge. Inspired by English as a natural language, we investigate the linguistic structure that ensures reliable communication of information and draw parallels with data structures, understanding both as models of systems of interest. We conceptualize semantic interoperability as comprising terminological and propositional interoperability. The former includes ontological (i.e., same meaning) and referential (i.e., same referent/extension) interoperability and the latter schema (i.e., same data schema) and logical (i.e., same logical framework) interoperability. Since no best ontology and no best data schema exists, establishing semantic interoperability and FAIRness of data and metadata requires the provision of a comprehensive set of relevant ontological and referential entity mappings and schema crosswalks. We therefore propose appropriate additions to the FAIR Guiding Principles, leading to FAIR 2.0. Furthermore, achieving FAIRness of data requires the provision of FAIR services in addition to organizing data into FAIR Digital Objects. FAIR services include a terminology, a schema, and an operations service.

# Background

With the total volume of data[1] doubling every three years (1) and the scientific community experiencing a surge in publications with 7 million academic papers being published annually (2), it is clear that **we need to harness machine assistance** to avoid that this overwhelming amount of data and knowledge prevents us from gaining meaningful insights and from wasting resources on research that has already been done elsewhere. Currently, a significant portion of research data are still scattered across various repositories and databases, using different data structures and terminologies, challenging researchers not only in finding and accessing them, but also in integrating them with other data to make them interoperable across different repositories and databases, and finally in reusing them for their specific research interests.

With this in mind, it is essential to facilitate **machine-actionable data** in scientific research, so that machines assist researchers in identifying relevant data pertaining to a specific research question. Moreover, enhancing the machine-actionability of data could contribute to a solution to the reproducibility crisis in science (3) by making raw data available, findable, and usable (4).

As a response to the need for machine-actionable data, in 2016, the **FAIR Guiding Principles for data and metadata** were introduced, providing a framework to assess the extent to which data are **F**indable, **A**ccessible, **I**nteroperable, and **R**eusable for both machines and humans alike (5). These principles have gained increasing attention from the research, industry, and knowledge management tool development communities in recent years (5–10). Furthermore, stakeholders in science and research policy have recognized the significance of the FAIR Principles. The economic impact of FAIR research data was estimated by the European Union (EU) in 2018, revealing that the lack of FAIR data costs the EU economy at least 10.2 billion Euros annually. Taking into account the positive effects of FAIR data on data quality and machine-readability, an additional 16 billion Euros were estimated (11). Consequently, the High Level Expert Group of the [European Open Science Cloud](#) (EOSC) recommended the establishment of an Internet of FAIR Data and Services ([IFDS](#)) (12). The IFDS aims to enable data-rich institutions, research projects, and citizen-science initiatives to make their data accessible in accordance with the FAIR Guiding Principles, while retaining control over ethical, privacy, and legal aspects of their data (following Barend Mons' data visiting as opposed to data sharing (13)). Achieving this goal requires the provision of rich machine-actionable data, their organization into FAIR Digital Objects (FDOs) (14,15), each identifiable by a Globally Unique Persistent and Resolvable Identifier (GUPRI), and the development of suitable concepts and tools for human-readable interface outputs and search capabilities. Although progress has been made toward building the IFDS (see the [GO FAIR Initiative](#), [EOSC](#), and the [FAIR Digital Objects Forum](#)), the current state of the FAIRness of data in many data-rich institutions and companies is still far from ideal.

The increasing volume, velocity, variety, and complexity of data present significant challenges that traditional methods and techniques for handling, processing, analyzing, managing, storing, and retrieving data struggle to address effectively within a reasonable timeframe (16). However, **knowledge graphs**, in conjunction with **ontologies**, offer a promising technical solution for implementing the FAIR Guiding Principles, thanks to their transparent semantics, highly structured syntax, and standardized formats (17,18). Knowledge graphs represent instances, classes, and their relationships as resources with their own GUPRIs. These GUPRIs are employed to denote relationships between entities using the triple syntax of *Subject-Predicate-Object*. Each particular relationship is thus modeled as a structured set of three distinct elements (i.e., data points), taking the form of resources and literals. In contrast, relational databases model entity relationships between data table columns and not between individual data points. Consequently, knowledge graphs outperform relational databases in handling complex queries on densely connected data (19), which is often the case with research data (with the exception of highly dimensional numerical data requiring complex analytical operations where knowledge graphs may face scalability challenges (20)). Therefore, knowledge graphs are particularly well suited for FAIR research and development, as well as any task requiring efficient and detailed retrieval of data.

---

[1] In the following, we subsume metadata (data about data) and data and refer to both as "data".

Nonetheless, employing ontologies and knowledge graphs to document data does not in itself guarantee adherence to the FAIR Principles. Achieving FAIRness necessitates meeting additional requirements, such as consistent usage of the same semantic data model for identical types of data statements to ensure schema interoperability (21), as well as organizing data into FAIR Digital Objects (14,15). Moreover, knowledge graphs, being a relatively new concept and technology, introduce their own specific technical, conceptual, and societal challenges. This is evident in the somewhat ambiguous nature of the knowledge graph concept (17) and the lack of commonly accepted standards, given the diverse technical and conceptual incarnations ranging from labeled property graphs like [Neo4J](#) to approaches based on the Resource Description Framework (RDF), employing RDF triple stores and applications of description logic using the Web Ontology Language (OWL). Ontologies, on the other hand, can significantly differ in quality and applying them correctly and logically consistent is not straightforward and requires experience in semantics. Consequently, we need tools that quantitatively and transparently measure the FAIRness of research data.

In this context, it is important to note that the **FAIR Guiding Principles are not a data standard** themselves, but a list of criteria that must be followed to obtain FAIR data—they do not specify, how FAIRness must be achieved. However, the popularity of the Principles triggered and informed the development of various FAIRness assessment tools that now start to serve as practical benchmarks within different research communities. Especially **automated FAIRness assessment tools** such as [FAIR-Checker](#) **(22)**, [F-UJI](#) **(23,24)**, and [FAIR Evaluator](#) **(25)** are of particular importance as they provide **transparent workflows for evaluating FAIRness scores** and are likely contributing to the establishment of corresponding new data standards. These automatic tools, however, have in common that they are limited to assessing only the FAIRness of basic provenance (e.g., creator, creation date) and licensing data (i.e., copyright license), but do not assess the **FAIRness of domain-specific data**. This is understandable, as they assess the FAIRness against a defined set of relevant and well established vocabularies, which are easier to specify for this kind of metadata but not as straightforward for domain-specific data if the assessment tool should be a general and not a domain-specific solution.

In this paper, we assume that the demand for FAIR and machine-actionable data presupposes their **successful communication between machines and between humans and machines**. During such communication processes, preserving the meaning and reference of the message between sender and receiver is crucial, requiring both parties to share the same background knowledge, encompassing lexical competencies, syntax and grammar rules, and relevant contextual knowledge. Based on this assumption, we analyze the different aspects that affect the semantic interoperability of data and suggest corresponding additions to the FAIR Guiding Principles, i.e., **FAIR 2.0**. We also shed some light on why achieving semantic interoperability across data management systems is such a big challenge. Part of the reason is that semantic interoperability requirements go beyond simply mapping terms across different controlled vocabularies. We also have to consider what is required for establishing semantic interoperability of statements and sets of statements, and thus granularity levels of information coarser than the level of particular terms.

In the *Result* section, we first provide a definition of what we understand by machine-actionability, before we are drawing inspiration from natural languages like English and explore how semantic interoperability can be understood by going into some detail in analyzing the way terms and natural language statements convey meaning and information. Recognizing terms and statements as basic units of information, we investigate the linguistic structures that ensure reliable communication of information and draw parallels with the structures found in data schemata as the machine-actionable counterparts to statements, understanding both as models that model some system-of-interest. When reading this section, you may ask yourself how this relates to data, FAIRness, and machine-actionability, but we believe that this analysis yields insights that help us to better understand what is required for achieving semantic interoperability of terms and statements. As a result of this analysis, we outline a general framework for conceptualizing semantic interoperability that distinguishes terminological and propositional interoperability. Whereas terminological interoperability involves sharing the same meaning/intension as well as sharing the same reference/extension, propositional interoperability involves sharing the same data schema and logical formalism.

Whereas the discussion around the FAIR Guiding Principles primarily focuses on the FAIRness of basic provenance and licensing data, our main focus in this paper lies on the question of what the requirements are for establishing FAIRness across all kinds of data. In the *Discussion* section, we argue that there is no simple solution for establishing cross-domain FAIRness of data. It is neither possible to develop a single best ontology for all domains of research and for the various scientific frames of reference that researchers adopt. Nor is it possible to develop a set of best data schemata for all kinds of operations one wants to conduct on data. Thus, semantic interoperability issues will inevitably remain to exist in the future. We therefore suggest additions to the FAIR Guiding Principles, resulting in FAIR 2.0, which reflect, among other things, requirements on data that we identified in our general framework for semantic interoperability. We also argue that for achieving FAIRness of data, we need FAIR Services in addition to organizing data in FDOs, and we discuss that they include a Terminology Service, a Schema Service, and an Operations Service. Consequently, the interoperability and FAIRness of data depends not only on the availability of readily applicable operations on the data (i.e., machine-actionability), but also on the provision of a comprehensive set of relevant ontological and referential entity mappings and schema crosswalks for them.

> **Box 1 | Conventions**
>
> In this paper, we refer to FAIR knowledge graphs as machine-actionable semantic graphs that are utilized for the purpose of documenting, organizing, and representing lexical, assertional (e.g., empirical data), universal, and contingent statements. We, thus, understand knowledge graphs to consist of a combination of empirical data and general domain knowledge, distinguishing them from ontologies, which primarily contain general domain knowledge and lexical statements, but no empirical data. Throughout the paper, we use the term *triple* to denote a triple statement, and *statement* to refer to a natural language statement. Also, when we talk about *schemata*, we explicitly include schemata for statements and for collections of statements and not only schemata for individual triples.
>
> To ensure clarity, both in the text and in all figures, we represent resources using human-readable labels instead of GUPRIs. It is implicitly assumed that each property, instance, and class possesses its own GUPRI.

# Result

## Machine-Actionability

The concept of machine-actionability of data is central to this paper. We therefore want to clarify what we mean by it. Looking at the definition in Box 2, it is clear that **machine-actionability cannot be simplified as a mere Boolean property**. Instead, it exists on a spectrum, allowing for **degrees of machine-actionability**. Numerous operations can potentially be applied to a given set of data, and the ability to apply even a single operation would suffice to classify the data as machine-actionable. Consequently, specifying the set of operations that can be applied to the data, along with the corresponding tools or code (i.e., *dataset A is machine-actionable with respect to operation X using tool Y*), is more meaningful than simply labeling it as machine-actionable.

> **Box 2 | Machine-Readability, Machine-Interpretability, Machine-Actionability** (21,26)
>
> **Machine-readable** are those elements in bit-sequences that are clearly defined by structural specifications, such as data formats like CSV, JSON, or XML, or resources and literals in the Resource Description Framework (RDF).
>
> **Machine-interpretable** are those elements in bit-sequences that are **machine-readable** and can be related with semantic artefacts in a given context and therefore have a defined purpose, such as referencing defined and registered ontology terms that provide **meaning** to a resource in an RDF triple following the triple syntax of *Subject-Predicate-Object*.
>
> **Machine-actionable** are those elements in bit-sequences that are **machine-interpretable** and belong to a type of element for which **operations** have been specified in symbolic grammar, thus linking types of data statements to operations such as logical reasoning based on description logics for OWL-based data and other rule-based operations such as unit conversion or other data conversions.

It is worth noting that a machine reading a dataset could be considered an operation itself. Therefore, datasets documented in formats as PDF, XML, or even ASCII files could be considered machine-readable and, to some extent, already machine-actionable. Moreover, if a dataset is machine-readable, search operations can be performed on it as well, enabling the identification of specific elements through string matching, for example. The success of search operations serves as a measure of the findability of data. Machine-readable data can be found through string matching, while interpretable data can be found through their meaning, referent, or contextual information. Thus, data that are readable but not interpretable possess limited findability. Consequently, analog to machine-actionability, **findability cannot be characterized as a Boolean property**.

It is important to emphasize that the definition of machine-actionability we refer to, as outlined in Box 2, strictly depends on machine-interpretability. Consequently, machine-reading of a dataset and machine-searching based on string matching are not considered proper examples for operations that fulfill the requirements of machine-actionability.

## Interoperability

Interoperability of data and metadata, as we understand it, is directly dependent on machine-actionability, where datasets *A* and *B* are considered *X*-interoperable if a common operation *X* exists that can be applied to both. As this definition is based on our understanding of machine-actionability, it requires *A* and *B* to be machine-interpretable, and it excludes the operations of machine-reading and string-based machine-searching. This characterization goes beyond the more generic definition of interoperability as the ability to exchange data (27). To achieve this kind of interoperability in the context of the exchange of data and metadata between different systems, data and metadata must be transferred in a way that guarantees that they remain functional and processable in the target system, which, in turn, requires data and service standardization (28).

As a consequence of the dependence of interoperability on machine-actionability, **interoperability inherits from machine-actionability that it is not a Boolean property, but rather exists along a continuum**, determined by the range of operations that can be executed on a given type of data. Achieving interoperability also entails the capacity to identify the specific type of data within a given dataset that is amenable to processing through a particular operation conducted by a specific tool or application, and vice versa. For example, identifying all subgraphs in a knowledge graph that model measurement data to which a particular unit conversion operation can be applied.

Data are composed of terms that form statements. With terms, we here refer to strings of literals and symbols that identify or represent real-world kinds, concepts, individuals, or values. The weight measurement of an apple *X*, for instance, is composed of terms representing the individual apple *X*, its particular weight, the measured value and the unit, forming the statement *'Apple X has a weight of 212.45 grams'*. Both terms and statements play a crucial role in conveying semantic content and thus meaning, forming the basis for successfully communicating information. Consequently, the interoperability of terms and of statements between sender and receiver of information is essential for effective communication. Exchanging data between machines and between a machine and a human being is not only about guaranteeing their readability, but also involves processing to ensure their interpretability (i.e., receiver understands their meaning) and their actionability (i.e., receiver can use them in another context by applying specific operations to them).

Evidently, interoperability plays a crucial role in facilitating this communication process and is central to the realization of FAIR data. Without interoperability, the findability and reusability of data are limited, and without interoperability there is no machine-actionability. The significance of interoperability has also been duly acknowledged by the EOSC. In their **EOSC Interoperability Framework** (14)[2], in accordance with the FAIR Guiding Principles, they distinguish four layers of interoperability for scientific data management:

- **technical interoperability** oversees interoperation at the application level within an infrastructure (i.e., information technology systems must work with other information technology systems in implementation or access without any restrictions or with controlled access),
- **semantic interoperability** is concerned with achieving shared semantic understanding of data elements (i.e., contextual semantics related to common semantic resources),
- **organizational interoperability** focuses on harmonizing business processes (i.e., contextual processes related to common process resources), and
- **legal interoperability** guarantees cooperation among organizations operating under diverse legal frameworks, policies, and strategies (i.e., contextual licenses related to common license resources).

---

[2] which, in turn, was inspired by the European Interoperability Framework from the European Commission from 2017.

Alternative differentiations of interoperability have been suggested. Sadeghi *et al.* (28), for instance, distinguish the following three facets of interoperability (i.e., the *interoperability trilogy*) in relation to needs they identified based on an interoperability survey they conducted:

1. **Data interoperability**: the need to exchange data, which is challenged by heterogeneous data.
2. **Service interoperability**: the need to use one another's services, which is challenged by disparate APIs and services.
3. **System interoperability**: the need to systematically manage the infrastructures for distributing, discovering, sharing, and exchanging artefacts, which is challenged by historical constraints and the lack of common practices to develop modular systems and the lack of supporting tools for modular software engineering.

In what follows, we focus on **semantic interoperability** and the facet of **data interoperability**.

## Semantic interoperability and what natural languages like English can teach us

The EOSC Interoperability Framework characterizes semantic interoperability as a requirement for enabling machine-actionability between information systems, and it is achieved *"when the information transferred has, in its communicated form, all of the meaning required for the receiving system to interpret it correctly"* (p. 11).

To understand what semantic interoperability implies at a conceptual level, it is helpful to consider how we as humans communicate meaning (i.e., semantic content) in a natural language such as English, using terms and statements as the basic units of information carrying meaning. And when we talk about communication, we mean the attempt to create the same cognitive representation of information in the receiver as is present in the sender.

Although the following section may seem very academic at first reading, and you may wonder what this has to do with data interoperability and the FAIR Guiding Principles, we believe that it provides the background necessary to better understand the complexity of semantic interoperability and to draw practical consequences for structuring data to achieve FAIRness—not only of basic provenance data but also of domain-specific data. In the subsequent sections, we then draw parallels between the structure of natural language statements and data structures by understanding both as models of some system-of-interest and outline a general framework for conceptualizing semantic interoperability, which is grounded in the observations we make in the following section.

**Requirements for successfully communicating terms and statements**

What is needed to communicate terms and statements efficiently and reliably? For successful communication between humans, both the sender and the receiver of the information need to share the same relevant background knowledge. Let us first take a look at what is needed for a natural language term to be readable, interpretable, and actionable before we turn to a natural language statement.

**Interoperability of terms**

We can say that a **term is readable** if it consists of a **sequence of characters** that **can be assigned to sounds**[3]. Following this notion, '*EGrzjEZhsmtrjE*' would be a readable term for most English readers, although the term would likely be meaningless to them, whereas '*Це можна прочитати*' is not, because it uses the Cyrillic alphabet which most English readers are not familiar with. Following this notion of readability, sender and receiver must use the **same set of characters** (i.e., the same alphabet) and **agree on an order and thus reading direction** for communicating terms.

The readability of a term can be contrasted with its interpretability. We call a **term interpretable**, if it consists of a sequence of characters that are **readable** and that **can be assigned to a specific meaning**. Following this notion, '*EGrzjEZhsmtrjE*' is readable but not interpretable, whereas the term '*tree*' is both readable and interpretable for most English readers. Typical examples of readable but not universally interpretable terms are acronyms such as '*RDF*'. Therefore, the full meaning of an acronym is usually introduced before the acronym is used in communication. Interpretability of a term is achieved if sender and receiver share the same **inferential lexical competence** (29) in the form of knowledge about the **meaning** of the term (i.e., its intension). Inferential lexical competence refers to the ability to understand and make inferences about the meaning of words and phrases based on contextual cues and linguistic knowledge. **Ontological definitions** that answer the question '*What is it?*' provide such knowledge (30). Sender and receiver must agree on **proper names** to refer to the same individual entities (i.e., particulars), **general terms** or **kind terms** to the same set of individuals that meet the defining properties of the terms, sometimes also called their extensions, and **verbs** or **predicates** to refer to the same specific types of actions or attributes. In other words, sender and receiver must agree on a **common terminology**.

A **term is actionable**, if it consists of a sequence of characters that is **interpretable** and to which the operations '**designation**' (i.e., the object is given, and the matching term has to be found) and '**recognition**' (i.e., the term is given, and the matching object has to be identified) can be applied. These operations require the sender and receiver to share the same **referential lexical competencies** (29) and thus **diagnostic knowledge** about the **reference** of a given term (i.e., its extension). The referent or extension of a term is thus the (real) entity to which the term refers. A proper name refers to an individual entity (e.g., the referent of the term *Earth* is the planet Earth) and a general term or kind term to a set of individuals, i.e., the instances of that term (e.g., the extension of the term *planet* are all planets that have existed and will exist). Referential lexical competence is the ability of a language user to understand and use words in a manner that accurately refers to objects, concepts, or phenomena in the world. Diagnostic knowledge is often communicated in the form of method-dependent recognition criteria, images, or exemplars that answer the question '*How does it look, how to recognize or identify it?*' (30), enabling the receiver to use the term correctly in designation or recognition tasks[4]. Referential lexical competencies are thus needed to use a term correctly in different contexts. They are essential for the **human-actionability of terms**.

---

[3] For the sake of simplicity, we will limit ourselves here to terms and not other symbols, such as emojis.

[4] Unfortunately, many ontologies only provide ontological definitions and no recognition criteria. For example, the term '*cell nucleus*' (FMA:63840) of the Foundational Model of Anatomy ontology (31) is defined as "*Organelle which has as its direct parts a nuclear membrane and nuclear matrix*". As we cannot see cell nuclei without using a microscope and staining a cell sample, the definition does not provide the practical diagnostic knowledge needed for designation and recognition tasks. The term therefore does not provide the information needed to build the referential lexical competencies required to make the term human-actionable.

The difference between inferential and referential lexical competence can be nicely illustrated by transferring these two concepts to the daily work of physicians: In many cases, a physician's ontological knowledge (c.f., inferential lexical competence) of bacteria in general is sufficient to know that a bacterial infection can be fought with antibiotics. However, having only general diagnostic knowledge of bacteria is not sufficient to reliably recognize (c.f., referential lexical competence) Lyme disease, for example, since symptoms of bacterial infections are often bacterium- and host-specific.

**Interoperability of statements**

Given that the meaning of a term is provided by its ontological definition that takes the form of one or more statements, one could argue that terms are only placeholders, i.e., surrogates, for their ontological definitions and thus for statements, and that only statements carry meaning. While this may be a stretch, the communication of meaning requires more than just terms, but also rules and structures to place several terms in a specific context, related by predicates.

**Statements carry meaning in addition to the meaning of the terms that compose them.** This becomes obvious when the positions of terms in a given sentence are changed, such as *"Peter travels from Berlin to Paris"* versus *"Peter travels from Paris to Berlin"*—the same set of terms carries two different meanings. Therefore, for the efficient and reliable communication of statements, the sender and the receiver of the information must share a set of rules and conventions for formulating sentences using terms.

But how is the meaning of a statement represented in the human brain? Understanding how the human brain creates cognitive representations of semantic content is an active area of research (32). The human brain is a highly interconnected complex system that is continually influenced by input signals from the body and the world, so that a given neuron does not function in isolation but is substantially influenced by its neural context (33). It is therefore not surprising that there is evidence for at least two forms of object knowledge representation in the human brain, supported by different brain systems, i.e., motor-sensory-derived and conceptual-cognition-derived object knowledge (34,35). Motor-sensory-derived object knowledge is object knowledge which is gained through direct sensory experiences and physical interactions with those objects. Conceptual-cognition-derived object knowledge is object knowledge which is acquired through cognitive processes such as categorization, abstract reasoning, and symbolic representation. It is therefore reasonable to assume that lexical concepts are stored as patterns of neural activity that are encoded in multiple areas of the brain, including taxonomic and distributional structures as well as experience-based representational structures that encode information about sensory-motor, affective, and other features of phenomenal experience (36). These findings suggest that the cognitive representation of the meaning of a statement is likely to take the form of a complex network of associations, analogous to a multidimensional mind-map. Thus, when attempting to communicate a statement, the sender must first translate this multidimensional mind-map into a one-dimensional sequence of terms, i.e., a sentence, and the receiver must translate it back into a multidimensional mind-map. These two translation steps are supported by a set of syntactic and grammatical conventions, shared by the sender and the receiver, for formulating sentences using terms.

According to the **predicate-argument-structure** of linguistics (37,38), the main verb of a statement, together with its auxiliaries, forms the statement's predicate. A predicate has a **valence** that determines the number and types of subject and objects, called **arguments**[5], that it requires to complete its meaning. Further objects, called **adjuncts**, may be additionally related to the predicate, but they are not necessary to complete the predicate's meaning. Adjuncts provide optional information, such as a time specification in a parthood statement—you can remove the time specification from a parthood statement and the statement still makes sense, whereas removing the object that designates the part would result in a nonsense parthood statement. Therefore, every statement has a subject phrase as one of its arguments, and can have one or more object phrases as further arguments and additional object phrases as adjuncts, depending on the underlying predicate.

In the syntax of a statement, each argument and adjunct of a predicate can be understood as having a specific **syntactic position** in the statement (i.e., position in a syntax tree), with each position having its own specific **semantic role** that the position expresses (Fig. 1B; see also Kipper et al.'s (39) verb lexicon [VerbNet](#), which extends Levin verb classes (40), and see thematic roles sensu (39)). The list of subject and object arguments of a predicate-argument-structure can be described by a list of **thematic labels**, each reflecting the semantic role of its syntactic position (e.g., OBJECT, QUALITY, VALUE, UNIT in Fig. 1), and the syntactic structure of a statement can be represented by an ordered sequence of such thematic labels (Fig. 1C). The thematic labels thus function as descriptors of semantic roles that are mapped onto positions in a given syntactic frame ((39), see also [PropBank](#) (41)).

Syntax trees, with their different syntactic positions and associated semantic roles, contribute substantially to the meaning of their sentences, and they are used to translate a web of ideas in the mind of a sender into a string of words that can be understood by a receiver, and translated back into a web of ideas (42). The clearer the semantic roles of the different positions are, the easier it is for a human being to understand the information.

In a sense, when considering that different syntax trees can share the same thematic label, which then interconnects them (i.e., the object of one sentence is the subject of another), we can understand graphs of interconnected syntax trees as the first knowledge graphs created by humans, and their use seems to be quite straightforward, providing a structure that is interoperable with human cognitive conditions.

With all this in mind, we can now state that a **statement is readable** if it consists of a **sequence of characters** that can be **assigned to terms and sounds**, with **rules indicating the end of a term and the end of a statement**. '*EGrzjEZ hsmtrjE.*' would be a readable statement, with the space and the period indicating the end of a term and the end of a statement, whereas '*EGrzjEZhsmtrjE*' is not a readable statement as it lacks this information—it would be read as a single term and not a statement consisting of several terms.

A **statement is interpretable**, if it consists of a sequence of terms that is **readable** and that, based on **conventions, can be assigned to a syntax tree with positions and semantic roles and thus to a particular meaning**. The meaning of a statement, however, goes beyond the meaning of the set of its terms. Someone can read "*The Waste Land*" from T.S. Eliot and understand every single word of it, but can still not understand its meaning due to its fragmented structure and references to multiple literary and cultural traditions. However, as a minimum requirement for the interoperability of a statement, sender and receiver must agree on a **common statement structure with positions for terms and a shared terminology**.

---

[5] Not to be confused with arguments in the sense of debates. In this context, an argument is a subject or an object that is required for a given predicate to form a semantically meaningful statement, whereas adjuncts can be removed from a statement and the statement is still meaningful.

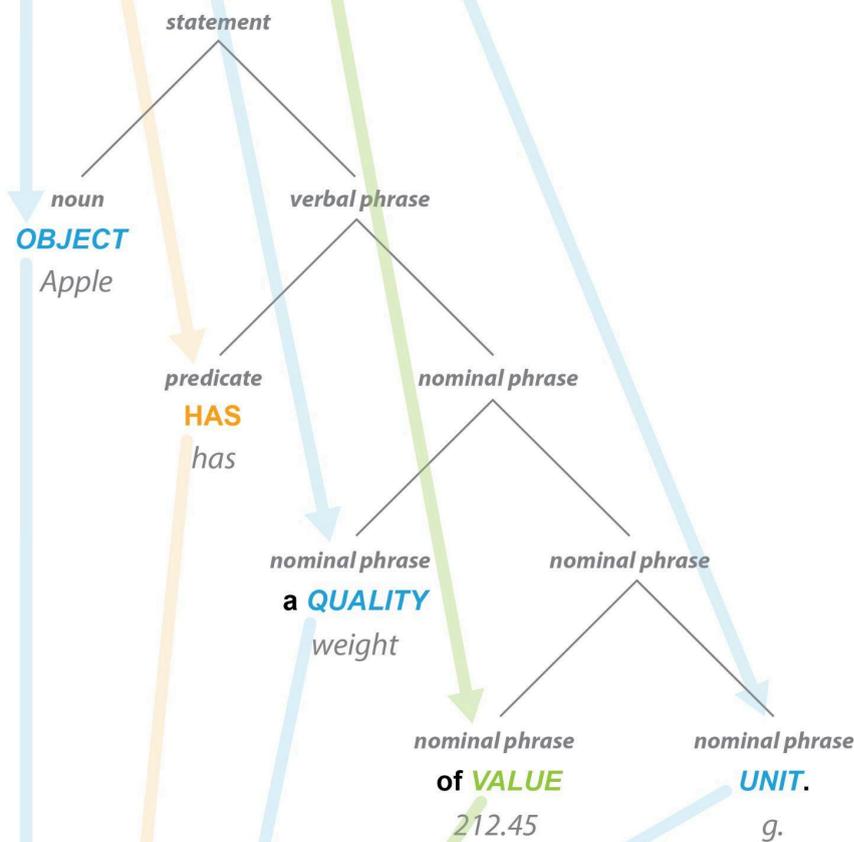

**Figure 1: Structure underlying a natural language statement.** The natural language statement in **A)** is structured by syntactic and grammatical conventions into syntactic positions of phrases of a syntax tree as shown in **B)** or a formalized statement as is shown in **C)**, where each position possesses a specific associated semantic role that can be described by a thematic label.

And finally, a **statement is actionable**, if it consists of a sequence of terms that can be **interpreted as a statement** and that **can be formed together with other statements into a meaningful narrative**. Sender and receiver **must agree on rules on how to refer to entities already mentioned before**, using for example pronouns such as '*he, she, it*'.

Since natural languages are highly expressive, **any given proposition (i.e., statement) can be usually expressed in more than one sentence**. Taking our example sentence from above, '*This apple has a weight of 212.45 grams*' (Fig. 1). The same information could also be communicated by the sentence '*The weight of this apple is 212.45 grams*' or '*212.45 grams is the weight of this apple*'. Although these three sentences have significantly different structures, by intuitively mapping the semantic roles associated with different syntactic positions across these sentences and by identifying that each position associated with the same role shares the same content, we immediately recognize that all three sentences communicate the same meaning (see Fig. 2). Thus, although different expressions may be used for the same information, usually no fundamental issues with semantic interoperability occur.

| **Natural Language Sentence** | **Formalized Statement** |
|---|---|
| *This apple has a weight of 212.45 grams.* | *OBJECT* **HAS** a *QUALITY* of *VALUE UNIT*. |
| *The weight of this apple is 212.45 grams.* | The *QUALITY* of *OBJECT* **IS** *VALUE UNIT*. |
| *212.45 grams is the weight of this apple.* | *VALUE UNIT* **IS** the *QUALITY* of *OBJECT*. |

**Figure 2: Different expressions of the same proposition. Left)** Three different expressions (i.e., natural language sentences) of the same proposition, each identical in their meaning. **Right)** The formalized statement for each natural language, with each syntactic position represented by its associated thematic role. The alignment of positions that share the same thematic role across the three statements is indicated by arrows.

To sum it up, whenever information needs to be communicated efficiently and reliably, the sender and receiver of the information must be able to identify terms and statements as the basic units of information in a message, recognizing where they begin and where they end for their **readability**. Moreover, they also must share the same inferential lexical competencies regarding the terms used in their communication and the same syntactic and grammatical conventions for formulating sentences with these terms, resulting in the same syntax tree in both sender and receiver for the **interpretability** and thus for communicating the **meaning** of the statement. Finally, sender and receiver must also share the same referential lexical competencies regarding the terms used in their communication to be able to correctly designate and recognize their referents, and they must share the same conventions for correctly placing statements in a **context** for their **actionability**.

**Parallels between the structure of natural language statements and data schemata**
The parallels between the structure underlying natural language statements about research findings (e.g., empirical observations, hypotheses, method descriptions) and corresponding data structures and their associated schemata become clear when considering that both are models representing a specific system-of-interest.

A model can be understood as information on something (i.e., meaning) created by someone (i.e., sender) for somebody (i.e., receiver) for some purpose (i.e., usage context) (43). The purpose of a model thereby is its use in place of the system it models—any answer that the model gives should be the same as what the system-of-interest would provide, restricting the model to those properties of the system that are relevant for the purpose (44). A model has to possess the following three features (45):

1. **Mapping feature**: a model is based on a system-of-interest, which it attempts to model;
2. **Reduction feature**: a model only reflects a relevant selection of the properties of its system-of-interest—no abstraction, no model;
3. **Pragmatic feature**: a model needs to be usable in place of its system-of-interest with respect to a specific purpose.

When we characterize models like this, we can understand both **syntax trees with semantic roles** and **data structures as models**. Token models can be distinguished from type models (44).

A **token model** (also called *snapshot model*, *representation model*, or *instance model*) models individual properties of elements of its system-of-interest and is thus based on a one-to-one correspondence between the model and the system, representing the system's individual attribute values such as the weight of a particular apple. **Token models thus model instances and their relations**. As a consequence, the creation of a token model involves only **projection** (i.e., choice of properties to be modelled) and **translation** (i.e., translating the properties to model elements). The sentence in Figure 1A and the table and graph representation of the same information in Figure 3B,C are examples for token models. The elements in a token model designate the corresponding elements of the system—here, a particular apple and its particular weight. As a consequence, different token models of the same system-of-interest that model the same system properties relate to each other through a transitive token-model-of relationship that can be ordered to chains of designators, each linearly designating its corresponding element across all token models, ultimately designating the corresponding element in the system-of-interest (44).

A **type model** (also called *schema model* or *universal model*), on the other hand, captures general aspects of a system-of-interest through **classification** of its properties. Taking a look at the relation between a sentence (Fig. 1A), its corresponding syntax tree (Fig. 1B) and formalized statement (Fig. 1C), we see that the semantic role of a given syntactic position can be obtained by classifying the object or subject instance of the sentence to a specific type or class (*'this apple'* → class *'apple'*), which in turn can be generalized to a corresponding semantic role (class *'apple'* → *'OBJECT'* role). Creating a type model thus involves classification in addition to projection and translation. Formalized statements with syntactic positions and their associated semantic roles (Figure 1C) are examples of type models. By classifying the entities allowed in a certain subject and object position, a sentence (i.e., token model) turns into a formalized statement type (i.e., type model). Graph patterns in the form of shapes or tables in a relational database are also type models, where the constraints of a specific slot or column specifies the class of allowed instances.

Instantiating a type model produces a corresponding token model. This allows us, in turn, to validate a token model against its corresponding type model. We can say that the sentence *'This apple has a weight of 212.45 grams'* is a token model that instantiates the type model *'OBJECT HAS a QUALITY of VALUE UNIT'* against which it can be validated. Applying different type models for modeling a given proposition, on the other hand, results in the creation of different token models of that same proposition (cf. Fig. 2).

According to (44), some type models are metamodels. A **metamodel** is a model of a model. In addition to projection, translation, and classification, metamodels involve **generalization**. As mentioned above, obtaining the semantic role of a syntactic position in a statement involves classification and generalization. Metamodels are more broadly applicable due to the generalization. Consequently, *'OBJECT HAS a QUALITY of VALUE UNIT'* is a metamodel, as it is a type model that generalizes the type model *'APPLE HAS a WEIGHT of VALUE UNIT'*. Formalized statements (cf. Fig. 1C) that model types of statements can be considered to be metamodels. A metamodel *A* is defined by being a type model that relates to another model *C* via a type model *B*, where the relation-chain from *A* via *B* to *C* is through two type-model-of relations. Most data structures are based on such metamodels, and when using language, we interpret sentences by using corresponding metamodels (see formalized statements in Fig. 1C and Fig.2 right). Any language, as such, can also be considered to be a metamodel, as it allows creating models. Thus, the structure used for documenting a datum (e.g., a row in a table or a subgraph in a knowledge graph) is a metamodel (e.g., English, CSV, or OWL) and the table structure or the graph pattern itself is also a metamodel (e.g., formal statement, CSV template, or [SHACL](#) shape).

Analog to the characterization of sentences as token models and their corresponding formalized statements as type models, we can understand a **proper name** to refer to an instance (i.e., individual) as its system-of-interest, by designating a **token model** of that system that specifies an ontological definition and operational recognition criteria that refer to properties of that system. A **kind term** or **general term**, in contrast, refers to a class or concept by designating a **type model** that specifies an ontological definition and operational recognition criteria that refer to properties of all instances of that class or concept.

Thus, if we understand the sentence *'This apple has a weight of 212.45 grams'* to be a natural language token model of a corresponding real apple, modeling the properties of that apple via corresponding syntactic positions, then we can conclude that each slot of a data model of the same real apple that models the same set of properties is a token model that relates to the corresponding syntactic position of the language model through a transitive token-model-of relationship. Consequently, comparing natural language and data representations of the same system-of-interest should be straightforward if they both model the same set of properties of the system.

Therefore, we can think of each datum as a somewhat formalized representation of a natural language statement, structured in such a way that it can be easily compared with statements of the same type, and easily read and operationalized by machines (cf. Fig. 3A with Fig. 3B,C). A datum is a token model that results from the instantiation of a data schema that is its corresponding type model. Both can be understood to be formalizations of a particular type of natural language token and type model that serve the purpose to support machine-actionability. In other words, data schemata are to machines what syntax trees are to humans—both define type models with positions and associated semantic roles for statements. When we compare data schemata with their corresponding natural language statements, we can thus see similarities between the structure of a sentence defined by the syntax and grammar of a natural language and the structure of a corresponding schema (Fig. 1,3). As discussed above, the **syntactic positions** of terms in a natural language sentence take on specific **semantic roles** and contribute significantly to the meaning of the statement.

For a data schema to have the same meaning as its corresponding natural language statement, it must, as a **minimum requirement**, provide a functionally and semantically similar structure with the same elements as the corresponding syntax tree: the schema must represent all relevant syntactic positions—in schemata often called **slots**—and their associated semantic roles in the form of **constraint specifications**, with terms and values populating the slots (see Fig. 3B,C). After all, humans need to be able to understand these data schemata, and need to be able to translate a given datum represented in a given data schema back into a natural language statement. If the column headers in tabular data structures are not properly characterizing the semantic roles associated with their respective syntactic positions (e.g., *'Location1'* and *'Location2'* instead of *'Departure Location'* and *'Destination Location'* for passenger transport data), humans will have difficulties interpreting them correctly. Data schemata should therefore be seen as attempts to translate the structure of natural language sentences into machine-actionable data structures.

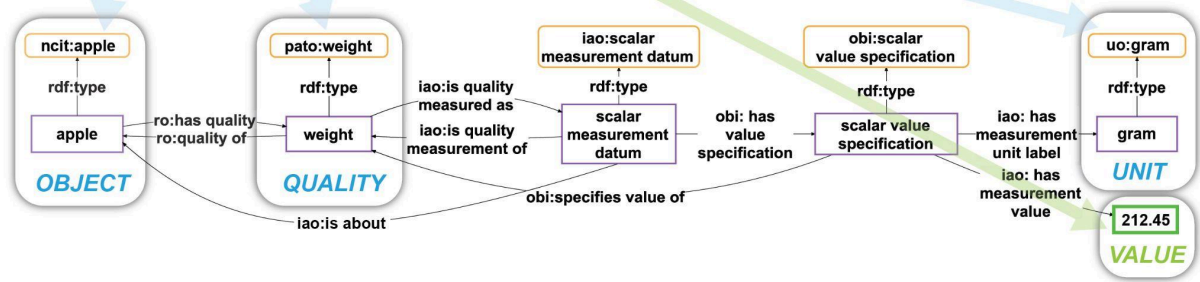

**Figure 3: Parallels between natural language statements and data schemata.** A data schema, both tabular as in **B)** and graph-based as in **C)**, must represent the same syntactic positions as its underlying natural language statement in **A)**. In a data schema, the positions take the form of slots, and each slot must specify its associated semantic role in the form of a constraint specification.

## A general framework for conceptualizing semantic interoperability

Looking at the interdependencies between data structures and natural language statements, we can distinguish different **causes for the lack of semantic interoperability of data**. And since we distinguish between terms and propositions (i.e., statements) as two different types of meaning-carrying entities when we communicate data, we can distinguish causes for the lack of terminological interoperability from causes for the lack of propositional interoperability.

**Terminological interoperability**

**Terminological interoperability** refers to the semantic interoperability of terms. Terms (i.e., entities) are used to identify or represent real-world concepts or instances and can be represented in data in the form of resources or literals. We can distinguish ontological and referential causes for the lack of terminological interoperability (21). When two given terms are compared semantically, they can either:

1. differ in both their meaning (i.e., their intension) and their referent (i.e., their extension) such as '*apple*' and '*car*',
2. differ only in their meaning but share the same referent, such as '*Morning Star*' and '*Evening Star*', which both refer to the planet Venus,
3. share the same meaning and the same referent/extension, or
4. differ in both their meaning and referent, but some ontological and referential overlap, closeness, or similarity relationship exists between them, such as '*apple*' and '*fruit*'.

If two terms share their meaning and their referent, they are **referentially and ontologically interoperable**, i.e., they are **strict synonyms** and can be used interchangeably. Since no two terms can share their meaning without also sharing their referent, ontological interoperability always implies referential interoperability, but not vice versa. Thus, if two terms have the same referent but not the same meaning, because controlled vocabularies may differ in their ontological commitments, their ontological interoperability is violated, but not necessarily their **referential interoperability**, since both terms may still be used to refer to the same entity. Two terms are referentially interoperable if they refer to the same set of (real world) entities, independent of whether they also share the same meaning. Consequently, **the set of ontologically interoperable terms is a subset of the set of referentially interoperable terms**.

For example, the COVID-19 Vocabulary Ontology (*COVoC*) defines '*viruses*' (NCBITaxon:10239) as a subclass of '*organism*' (OBI:0100026), while the Virus Infectious Diseases Ontology (*vido*) defines '*virus*' (NCBITaxon:10239) as a subclass of '*acellular self-replicating organic structure*' (IDO:0002000), and thus as an object that is not an organism (*vido* also reuses '*organism*' (OBI:0100026), but does not classify '*virus*' (NCBITaxon:10239) as one of its subclasses)–these two terms are therefore not ontologically interoperable, even though they have the same referent (i.e., the same extension)[6].

Researchers sometimes disagree on the classification of a given entity, or, when modelling the same system-of-interest, focus on different aspects of that system. As a consequence, they use different concepts to refer to the same thing. In all these cases, they thus agree on the referent but disagree on its ontological definition, i.e., the concept's meaning. Sometimes, researchers also change their mind due to new insights and change the classification of an entity and thus change the model. Pluto is a good example of the latter, which has been recently classified as a dwarf planet instead of a planet. We thus can distinguish $Pluto_{dwarf\ planet}$ and $Pluto_{planet}$, which are two different models that both refer to the same astronomical body from the solar system. Processes like this are rather common in empirical research, where some phenomenon is in need of an explanation and in the course of research, new theories emerge that improve our understanding of the phenomenon. Each new theory provides a new ontological definition, i.e., a new model, although they all share the same referent.

As far as terminological interoperability is concerned, we can therefore conclude that although ontological interoperability is preferred[7], **referential interoperability is the minimum requirement for the interoperability of terms**, since when we communicate information, we at least want to know that we are referring to the same (real) entities.

In addition to these two clearly demarcated cases with actionable consequences for terminological interoperability, we can recognize several intermediate relations between terms that neither share their meaning nor their referent, but nevertheless share some ontological overlap, closeness, or similarity. For instance, if the former is a special case of the latter, the terms can be used interchangeably in most contexts, or are just related. While this may represent useful information, it is not directly actionable in the context of terminological interoperability.

---

[6] And even the same identifier, since both ontologies have imported the '*Viruses*' class (NCBITaxon:10239) from NCBITaxon.

[7] However, in order to achieve ontological interoperability, the schemata used for the statements in the class axioms (i.e. their ontological definitions) of the terms to be mapped must also be semantically interoperable. If the schemata underlying the axioms differ, schema crosswalks must be specified—see propositional interoperability below. If the additional terms used in class axioms differ across class axioms, entity mappings must also be specified for them, provided they have the same meaning and the same referent. This is necessary because the meaning of a term is conveyed by its definition, which is a statement in its own right, with all the resulting consequences for propositional interoperability. Unfortunately, however, this propositional aspect of terminological interoperability is often overlooked.

In the context of knowledge graphs and ontologies, if two terms share the same meaning and the same referent despite having different GUPRIs, we can express their terminological interoperability by specifying a corresponding entity mapping using the '*same as*' (owl:sameAs[8]; skos:exactMatch) property. If two terms have only the same referent but not the same meaning, we can express their referential interoperability by specifying a corresponding entity mapping using the '*equivalent class*' (owl:equivalentClass) property. **We can thus distinguish between ontological (i.e., same-as) and referential (i.e., equivalent-class) entity mappings**. Both types of mappings are **homogeneous definition mappings** (47), where there is only one vocabulary element to be mapped on the left side, and several others on the right side of the definition that do not need to be mapped. Additionally, entity mappings can document further relationships, which can be helpful in solving interoperability issues (see Table 1).

**Table 1: A table of properties that can be used to specify relations between two terms in an entity mapping and their applications.** The property indicated by * is a property we suggest for specifying referential entity mappings in case owl:equivalentClass should not be used.

| entity mapping relation | application |
| --- | --- |
| owl:sameAs; skos:exactMatch | A transitive and symmetrical relation between two resources (i.e., terms) that are either concepts referring to an individual or instance (i.e., proper name) or to a class (i.e., general term or kind term), where both terms share the same meaning (i.e., intension) and referent (i.e., extension). We suggest using them for indicating ontological interoperability between concepts.<br>*Example: uberon:multicellularOrganism (UBERON:0000468) and ocimido:multicellularOrganism (OCIMIDO:00467).* |
| owl:equivalentClass; *new:referentialMatch* | A transitive and symmetrical relation between two resources (i.e., terms) that are either concepts referring to an individual or instance (i.e., proper name) or to a class (i.e., general term or kind term), where both terms share the same referent (i.e., extension) but not necessarily also the same meaning (i.e., intension). We suggest using them for indicating referential interoperability between concepts.<br>*Example: uberon:multicellularOrganism (UBERON:0000468) and caro:multicellularOrganism (CARO:0000012).* |
| owl:equivalentProperty | A transitive and symmetrical relation between two properties, where both properties share the same extension but not necessarily also the same meaning. We suggest using them for indicating referential interoperability between properties.<br>*Example: dcat:hasVersion and pav:hasVersion.* |
| rdfs:subClassOf | A transitive relation between two classes, where the domain (i.e., subject) specifies the parent class and the range (i.e., object) the subclass.<br>*Example:* foodon:animal (FOODON:00003004) and *uberon:multicellularOrganism (UBERON:0000468).* |
| rdfs:subPropertyOf | A transitive relation between two properties, where the domain (i.e., subject) specifies the parent property and the range (i.e., object) the subproperty.<br>*Example: ro:hasComponent (RO:0002180) and bfo:hasPart (BFO:0000051).* |
| skos:closeMatch | A non-transitive, symmetrical relation between two concepts (either instances or classes), indicating that they are sufficiently similar to be used interchangeably in some information retrieval applications. This relation may be of value to human readers, but it is not defined by formal semantics and thus cannot be used by machines in any |

---

[8] This is straightforward for mapping across terms that refer to individuals, and indicates that the mapped resources actually refer to the same individual. In OWL Full, however, where classes can be treated as instances of (meta)classes, it can also be used to specify that the two mapped classes not only have the same class extension and thus the same reference (which can be expressed using owl:equivalentClass) but also the same intensional meaning and thus the same ontological definition (46).

| | meaningful sense. |
|---|---|
| skos:relatedMatch | A non-transitive, symmetrical relation between two concepts (either instances or classes), indicating an associative relation between them. This relation may be of value to human readers, but it is not defined by formal semantics and thus cannot be used by machines in any meaningful sense. |
| skos:broadMatch | A non-transitive, hierarchical relation between two concepts (either instances or classes), indicating that the object is broader than the subject. This relation may be of value to human readers, but it is not defined by formal semantics and thus cannot be used by machines in any meaningful sense. |
| skos:narrowMatch | The inverse relation of skos:broadMatch. |

**Propositional interoperability**

Machine-actionable information necessarily takes the form of statements (i.e., propositions)—in knowledge graphs and ontologies either as ABox semantic instance-graphs and thus assertional statements or as TBox semantic graphs in the form of class axioms and thus as universal statements.

**Propositional interoperability** refers to the semantic interoperability of statements. We can distinguish logical and schematic causes for the lack of propositional interoperability (21). Data and metadata statements are **logically interoperable** if they have been modeled on the basis of the same logical framework (e.g., OWL2-DL), so that one can reason over them using appropriate reasoners (e.g., Pellet (48)). It is important to realize that logical interoperability and logical consistency are two different data characteristics: logical inter**operability** only depends on whether **reasoning operations** can be applied to the data. It does not necessarily include that the data are also logically consistent. When we talk about logically interoperable terms, we are actually referring to the ontological definitions of these terms and thus to their class axioms, which are universal statements that are logically interoperable if they are represented using the same logical framework.

**Schema interoperability** is achieved when statements of the same type are documented using the same data schema. If statements of the same type were represented using different schemata, corresponding data would no longer be interoperable. In such cases, one would have to specify schema crosswalks (i.e., schema mappings) by aligning slots (e.g., syntactic positions, tables or columns in tabular data formats) that share the same constraint specification (i.e., the same semantic role) across different schemata modeling the same type of statement, in order to regain schema interoperability (see Fig. 4). This can be compared to how we recognize different natural language sentences as different expressions of the same proposition (cf. Fig. 2). If the schemata use different vocabularies to populate their slots (i.e., the constraint specifications of their slots refer to different vocabularies, the values in the cells of tabular data formats use different vocabularies), then corresponding entity mappings must be included in the crosswalk to ensure terminological interoperability (see red bordered slots in Fig. 4). Consequently, we can distinguish **ontological and referential schema crosswalks** as two poles of a continuum. A schema crosswalk is a set of rules that specifies how data elements or attributes and thus slots from one schema and format can be aligned to the equivalent slots in another schema and format, with the constraint specification of aligned slots being mapped to each other using entity mappings. Schema crosswalks thus go beyond entity mappings, as they do not merely map one individual term to another but align slots for terms based on their semantic roles. Consequently, we need specific minimum metadata standards for schema crosswalks in addition to standards such as the SSSOM (see Box 3) for entity mappings.

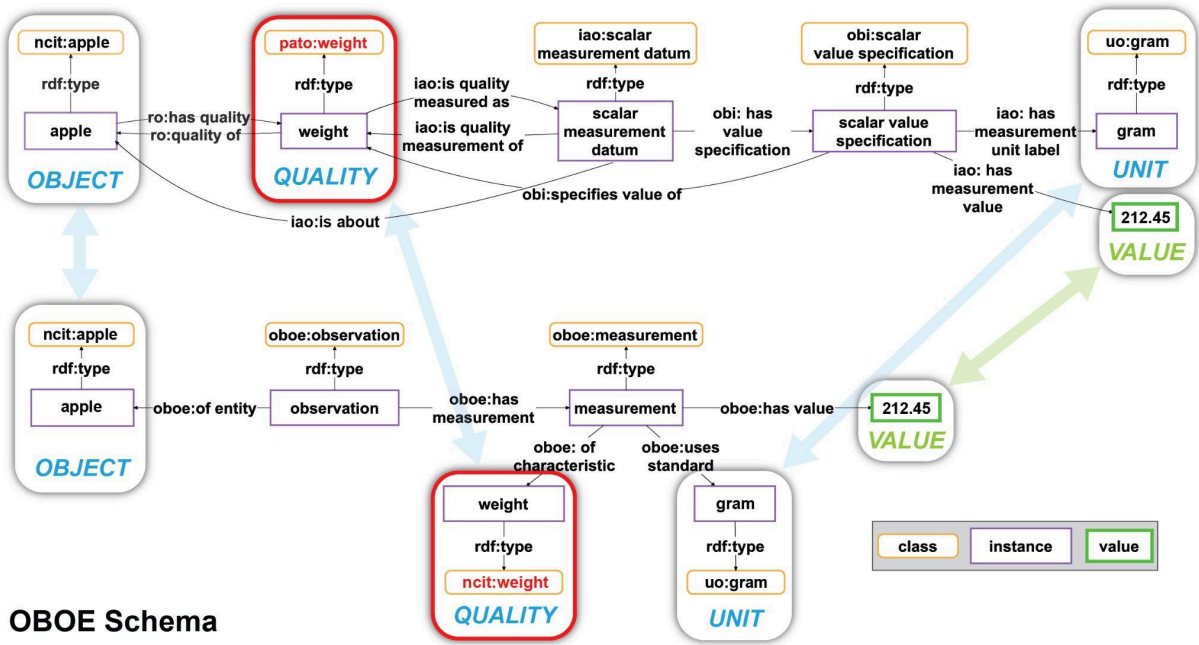

**Figure 4: Crosswalk from one schema to another for a weight measurement statement.** The same weight measurement statement is modeled using two different schemata. **Top**: The weight measurement according to the schema of the Ontology for Biomedical Investigations (OBI) (49) of the Open Biological and Biomedical Ontology (OBO) Foundry, which is often used in the biomedical domain. **Bottom**: The same weight measurement according to the schema of the Extensible Observation Ontology (OBOE), which is often used in the ecology community. The arrows indicate the alignment of slots that share the same constraint specification, i.e., the same semantic role. The corresponding semantic roles include the *OBJECT*, the *QUALITY*, and the *VALUE* that has been measured together with its *UNIT*. The slots carry the information that actually conveys the meaning of the weight measurement statement to a human reader. Blue arrows indicate slots with resources as values, and green arrows those with literals. Slots with red borders indicate problems with terminological interoperability: OBO uses an instance of the class 'pato:weight', while OBOE, in this example, uses an instance of the class 'ncit:weight'. However, since 'pato:weight' and 'ncit:weight' are synonymous terms and can therefore be mapped, the problem can be resolved with a corresponding entity mapping.

In other words, to achieve schema interoperability between two given statements, the subject, predicate, and object slots of their data schemata need to be aligned, and their terms need to be mapped across controlled vocabularies. To do this, the schemata must first be formally specified, e.g., in the form of graph patterns specified as SHACL shapes (see Box 3). Shapes that share the same statement-type referent then need to be aligned and mapped. These are **ontology pattern alignments** (i.e., **type model** (i.e., TBox) alignments) (47) or **token model** (i.e., ABox) **alignments**, where several vocabulary elements must first be aligned and then mapped in schema crosswalks. If two schemata share the same constraints across all of their slots (i.e., any statement expressed in schema *A* can be expressed in schema *B* using the same set of terms without violating their slot-constraints), the corresponding crosswalk can specify them as ontologically identical (analogue to entity mappings, via owl:sameAs or skos:exactMatch). All other crosswalks can be specified as referentially identical (via owl:equivalentClass).

As far as propositional interoperability is concerned, we can therefore conclude that although a combination of logical interoperability and ontological schema interoperability is preferred, for the same reasons as for terms, **schema interoperability using referential schema crosswalks is the minimum requirement for the interoperability of statements**.

Regarding the concept of *model* as discussed above, entity mappings and schema crosswalks both represent information on the transformation of a source to a target model and are thus models of model-to-model transformations with the purpose to automate a translation process (on transformations as model see (44)). Such a transformation only translates those properties (i.e., syntax positions with associated semantic roles or slots with constraints) of the source to the target model that are relevant for the target model's purpose. In the case of entity mappings, they model the transformation between token models (i.e., between two proper names) and between type models (i.e., between two kind or general terms), and in the case of schema crosswalks between two corresponding type models.

In summary, the interoperability of data statements does not only depend on the number of applicable operations and thus on machine-actionability, but also on the completeness of the ontological and referential entity mappings and schema crosswalks available that are relevant to the statements.

> **Box 3 | Existing Work on Entity Mappings and Schema Crosswalks**
>
> [SHACL](#) (50) and [ShEx](#) (51) are shape constraint languages for describing RDF graph structures (i.e., shapes) that identify predicates and their associated cardinalities and datatypes. Shapes can be used for communicating data structures, creating, integrating, or validating graphs, and generating user interface forms and code.
>
> RDF Mapping Language (RML) (52) is an extension of the W3C recommendation R2RML that provides a generic mapping language used to express rules for the bidirectional mapping between data in heterogeneous structures and serializations, including JSON, XML, CSV, and SQL databases and the RDF data model. RML maps are RDF graphs, with classes and properties defined in the RML ontology and constraints defined in corresponding SHACL shapes. RML allows data transformations, computations, and filtering without requiring changes to the underlying data source and thus enables schema crosswalks. Tools and resources to process, edit, and validate RML mapping rules are provided via the rml.io platform.
>
> SDM-RDFizer (53) is an interpreter of RML that implements novel algorithms to process RML mappings faster, allowing to scale up to complex scenarios where data is not only broad but has a high-duplication rate.
>
> [LinkML](#) (54) is a general purpose modeling language for defining schemata and data dictionaries in human reader friendly YAML syntax. LinkML provides generators to translate LinkML schemata automatically into other commonly used schema languages such as JSON-Schema, ShEx, RDF, OWL, GraphQL, and SQL DDL, as well as Python dataclasses and an HTML representation for human users. It is designed to easily map each schema item (classes, slots, datatypes and enumerations) to terms from existing terminologies via their GUPRI, thus facilitating semantic interoperability. The LinkML framework also provides a variety of different tools with varying maturity, to address additional use cases such as schema based data curation and transformation, inferring schemata, building schemata from spreadsheets, enabling schema crosswalks based on declarative mapping rules, or using Large Language Models to generate LinkML schema conform data from unstructured text. This broad functionality of LinkML fosters interoperability in many different contexts.
>
> The [Simple Standard for Sharing Ontological Mappings](#) (**SSSOM**) (55–57) has been developed as a standardized data model for the exchange of entity mappings. It primarily addresses two needs: defining an easy-to-use format for exchange and a detailed vocabulary for provenance and mapping justifications. Mapping justifications describe the processes by which a mapping was determined, such as lexical matching, manual curation, or semantic matching. SSSOM is modeled in LinkML, has been integrated into ontology curation tools such as the Ontology Development Kit, the Ontology Access Kit, ROBOT, a Python library and command line interface, and it is also proposed as the standard within the FAIR-IMPACT project of the EOSC (58).
>
> The [InteroperAble Descriptions of Observable Property Terminology](#) (**I-ADOPT** (59)) is an interoperability framework for representing observable properties, developed by the Research Data Alliance ([RDA](#)). It is based on an ontology designed to enhance interoperability between complex observable properties, each component of the model (e.g. property, entity, constraint) can be described using existing terminologies(e.g., ontologies, taxonomies, controlled vocabularies). The framework encompasses a [repository](#) of proposed terminologies to compose semantic variable descriptions. Variables are understood to be descriptions of something observed or mathematically derived. I-ADOPT does not cover concepts such as units, instruments, methods, and geographic location information, but is confined to the description of the variable itself. It provides templates, i.e., Variable Design Patterns ([VDP](#)s), that are similar to Ontology Design Patterns and provide schemata for specific types of variables.
>
> The [Cross-Domain Interoperability Framework](#) (**CDIF**) is currently being developed to become a set of guidelines for using existing standards, such as persistent identifiers like [DOI](#), [ORCID](#), and [ROR-ID](#), schemata and models such as [Dublin Core](#), [SKOS](#), [OWL](#), [schema.org](#), [DCAT](#), [PROV-O](#), and I-ADOPT, in a coordinated way, to ensure a degree of FAIR exchange in as automated a fashion as possible. Building upon the [FAIR Digital Object Framework (FDOF)](#), CDIF aims to become a *lingua franca* for FAIR data exchange.

## Discussion

## How to achieve semantic interoperability: What makes a term a good term and a schema a good schema?

One might think that we could achieve semantic interoperability by *just* developing a common standard, format, and schema for each type of entity and each type of statement to which every data and metadata provider and consumer agrees on. This standard and format must be as rich as the constituent system models. In other words, the goal would be to achieve full semantic harmonization across all data and metadata sources, requiring convincing all stakeholders to agree on a common standard. This **integrated interoperability** (60) approach to achieving interoperability is presumably one of the reasons for the proliferation of standards that we see in so many areas.

For terms, this approach would require agreeing on a universal terminology. This, however, is not feasible. Even when ignoring societal, psychological, and historical factors, different research communities often apply different **frames of reference** and thus emphasize different aspects of a given system-of-interest they want to model, resulting in the **need for different terms for the same type of system**, inevitably resulting in issues with ontological interoperability. For instance, for some studies and experiments, an ontology that is based on quantum mechanics, in which an electron is both a particle and a wave, would not be adequate for modeling the aspects of reality relevant to the experiment and, instead, an ontology that is based on Newtonian physics, in which an electron is only a particle, would be preferable. Consequently, there is sometimes a **legitimate demand for more than one term for a particular system-of-interest**—a fact that is in direct opposition to the goal of a universal terminology. If we understand a term as a model of a system-of-interest that has been developed with a specific purpose in mind and if we consider that this always involves a certain degree of abstraction and reduction, it becomes clear that for some systems we need more than one model. Especially if the system is rich, i.e., if it has multiple sets of properties, each of which is relevant in a particular context.

For schemata, the situation is similar. A good schema for a data statement must cover all the information that needs to be documented, stored, and represented for the corresponding type of statement. However, beyond that, there are many other criteria for evaluating schemata. Most of these relate to the **different operations** one wants to perform on the data, and thus on the underlying purpose of the schema as a model of a system-of-interest. Each operation likely has different requirements on the schema in terms of performance optimization. Moreover, format requirements from corresponding tools and thus demands of **fitness-for-use,** allowing direct use of data and metadata, also influence the overall **degree of machine-actionability** of the data, and thus the choice of a data schema. Optimizing the findability of measurement data, for instance, likely requires a different data schema than optimizing reasoning over them. A given schema must therefore be evaluated in terms of the operations to be performed and the tools to be used on the data, often involving trade-offs between different operations with different priorities to achieve an overall optimum.

Therefore, although agreement on a universal terminology and a universal set of schemata would be a solution for semantic interoperability and machine-actionability of data across different research domains, this is unlikely to happen for the reasons discussed above, and **we have to think pragmatically** and emphasize the need for ontological and referential entity mappings for terminological interoperability and schema crosswalks for schema interoperability. Thus, 'integrated interoperability' is not a feasible approach to achieving interoperability in the FAIR sense.

Alternatively, the **unified interoperability** approach has been suggested (60), according to which semantic interoperability is achieved by agreeing on a common meta-level structure for establishing comparable representations of diverse formats and schemata via super metamodels. If a **super metamodel** attained widespread adoption as a reference, various formats and schemata could be mapped to it, allowing data conversion across these formats and schemata using the super metamodel as mediating structure, achieving semantic interoperability across all formats and schemata (see, e.g., *reference term* and *reference schema* in Fig. 5). Examples of this approach are: the Cross Domain Interoperability Framework that is intended to be used as a *lingua franca* for data sharing across domain boundaries, LinkML as a *lingua franca* for different schema modeling languages, allowing the export of a LinkML schema to other representations, and I-ADOPT that is aimed at unifying the semantic description of research variables (see Box 3).

The **federated interoperability** approach, on the other hand, assumes that data formats and schemata have to be adapted dynamically instead of having a predefined super metamodel. A **syntactic variant** of federated interoperability has been suggested, in which multiple point-to-point conversions are required that are achieved by specifying $n^2$ entity mappings and schema crosswalks for federating $n$ alternative terms and schemata. Ontology mappings typically take the form of such point-to-point entity mappings.

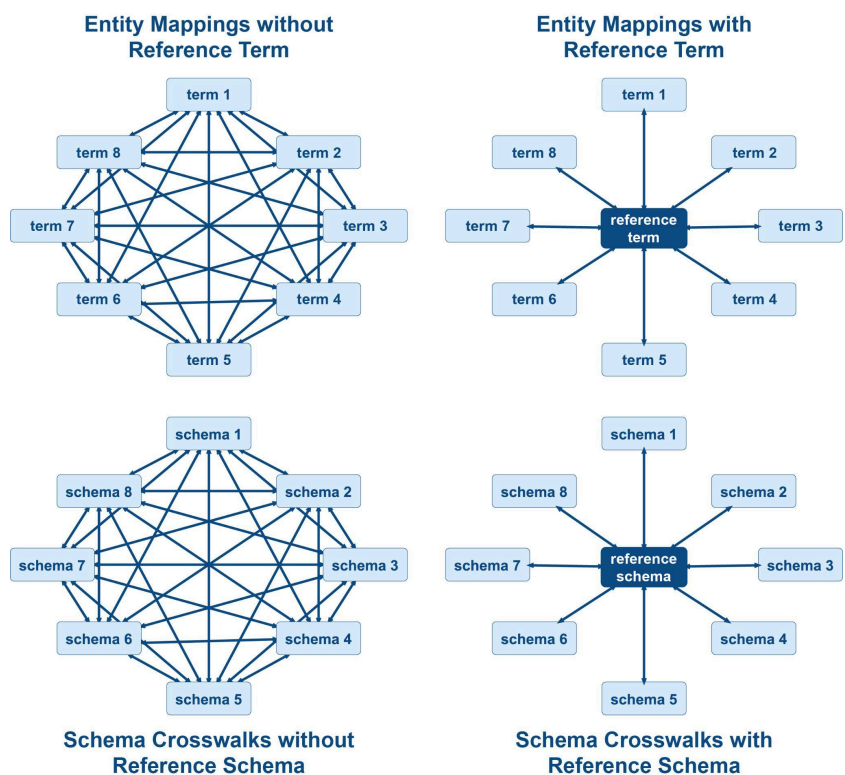

**Figure 5: Number of entity mappings and schema crosswalks required. Left)** 28 entity mappings and schema crosswalks respectively are required to achieve terminological and schematic interoperability between 8 different terms and schemata, because each possible pair of terms and schemata requires its own mappings and crosswalks. **Right)** With a reference term and a reference schema playing the role of an intermediary, the number of required mappings and schema crosswalks can be reduced to a minimum of only 8, which significantly reduces the effort required to establish terminological and schematic interoperability for the corresponding type of term and data statement.

Sadeghi *et al.* (28) suggest that the syntactic variant could be replaced by a **semantic variant** of federated interoperability that would provide meaning of data structures in a machine-interpretable way through the use of one or more ontologies. They argue that ontologies would provide agreed meaning of the concepts used in the data structures, resulting in a scalable approach, requiring only mappings to ontology resources, reducing the number of required mappings to *n* for federating *n* alternative terms, instead of $n^2$ as in the syntactic variant. However, if the semantic variant presupposes a semantic abstraction layer to which each system can map and crosswalk their data and artefacts, then this approach becomes equivalent with the unified interoperability approach, with the semantic abstraction layer taking the role of the super metamodel. Above, we already discussed why ontologies cannot serve this purpose. Moreover, the authors mention only entity mappings and do not explain how this approach achieves propositional interoperability.

## FAIR 2.0: Extending the FAIR Guiding Principles

From the above, we can conclude that achieving semantic interoperability is a complex task, and establishing the Internet of FAIR Data and Services (IFDS) is therefore a real challenge that requires more than merely organizing data into FAIR Digital Objects and specifying mappings across terms for achieving terminological interoperability. So far, discussions about semantic interoperability and corresponding tools and services have mainly revolved around terminological interoperability, focussing on entity mappings across various ontologies and controlled vocabularies. This is not surprising, considering that solutions for propositional interoperability must build on solutions for terminological interoperability. Moreover, whereas a term (in the broad sense, including literals) can be formally conceptualized in RDF as either an ontology term or a literal with associated datatype specification and thus as a clearly specified and demarcated entity, specifying and demarcating a data statement is not as straightforward. When modelled in RDF, a data statement can consist of one or more triples (e.g., Fig. 3C). Since we currently lack agreed upon semantic categories at granularity levels coarser than terms or single triples, it is not obvious how to conceptualize and clearly demarcate data statements. However, because information and meaning is communicated through such statements and not through individual terms, we need to find a solution for achieving propositional interoperability to obtain FAIR data and we need to understand statements (i.e., propositions) as a basic unit of information. And in a next step, we have to think about how to make collections of statements interoperable. We thus need, in addition to ontologies and entity mappings for terminological interoperability also something analogous for statement types[9] and collections of statements. It also has to be clarified what is required for statements and collections of statements to be FAIR.

We believe that the FAIR Guiding Principles need to be extended to cover the FAIRness of statements and collections of statements (see also Box 4). We therefore suggest the following additional sub-principles to be incorporated into the FAIR framework to achieve comprehensive semantic interoperability:

- **F5.1/I7.1**: The need to map terms with the same meaning and reference/extension to each other, effectively establishing entity mappings for terminological interoperability. These mappings must follow an established standard (e.g., SSSOM, see Box 3) and should differentiate ontological from referential and other types of entity mappings. By ensuring entity mappings across relevant vocabularies, datasets can be seamlessly communicated and information exchanged, fostering efficient data integration and analysis. This sub-principle relates to both Findability and Interoperability.

---

[9] Statement types are characterized by their underlying main verb or predicate. All statements about the weight of a particular object belong to the weight measurement statement type, and all statements about parthood relations between objects to the parthood statement type.

- **F5.2/I7.2**: The need to provide synonyms and language-specific labels for multiple languages for the terms that are used in data. This sub-principle relates to both Findability and Interoperability.
- **F6.1/I8.1**: Uniform data schemata must be maintained for statements of the same type, while referencing the schema's identifier in the statement's metadata. In some cases, (meta)data comprise specific collections of statements (e.g., material data sheets that characterize a specific type of material following an established standard). In such cases, the identifier must reference a schema that is the collection of the corresponding individual statement schemata. By adhering to this principle, propositional interoperability is improved, as statements with consistent schemata can be efficiently queried and processed. This sub-principle relates to both Findability and Interoperability.
- **F6.2/I8.2**: All data schemata relevant for statements of the same type are ideally aligned and mapped to each other in the form of schema crosswalks. By achieving propositional interoperability through schema crosswalks, datasets can effectively exchange information and ensure compatibility across various data representations. This sub-principle relates to both Findability and Interoperability.
- **F7./I6.**: Data must use a formalism that clearly distinguishes between lexical, assertional, contingent, prototypical, and universal statements (61,62). **Lexical statements** (*terminological statements* sensu (62)) are about **linguistic items** such as terms and comprise information such as the label, identifier, and the human-readable definition of a resource and the specification of its synonyms. In ontologies, lexical statements are usually documented using annotation properties. **Assertional statements** state **what is the case** (e.g., *This swan is white*). They are statements that are true for specific particulars. Empirical data are assertional statements. **Contingent statements** state **what can be the case** (e.g., *Swans can be white*). They are true for some instances of a specific type of entity. **Prototypical statements** are a subcategory of contingent statements and state **what is likely the case** (e.g., *Swans are typically white*). They are considered to be true as long as not the contrary is explicitly stated. **Universal statements** state **what is necessarily the case** (e.g., *All swans are white*). They are true for every instance of a specific type of entity.

    The two statements '*This swan is white*' and '*All swans are white*', despite using the same set of terms (e.g., *swan*='Cygnus' (NCBITaxon:8867); *white*='white' (PATO:0000323)), have a substantially different meaning. Distinguishing these different categories of statements is thus essential for human readers for correctly interpreting their meaning, and thus contribute to their propositional interoperability. Moreover, classifying data statements along these five categories also contributes to their findability. This sub-principle relates to both Findability and Interoperability.
- **A1.3**: To encompass organizational and legal interoperability, data must comply with existing data protection regulations, such as the General Data Protection Regulation ([GDPR](GDPR)). By adhering to relevant data protection regulations, organizations can facilitate secure and compliant data sharing, promoting seamless collaboration and data utilization. This sub-principle relates to Accessibility.
- **I4.**: Vocabularies used for documenting data and metadata must specify an ontological definition that is human-readable (not necessarily a machine-actionable class axiom), characterizing the ontological nature of a specific entity and should, where applicable, provide recognition criteria (i.e., an operational definition) specifying how to identify the entity to serve as FAIR vocabularies. Only by explicitly specifying ontological definitions and recognition criteria are terms made interpretable and actionable for human users, allowing them to not only understand their meaning but enable them to successfully apply them in designation (i.e., object given, matching term sought) and recognition tasks (i.e., term given, matching object sought). This sub-principle relates to Interoperability.

- **I5.**: The logical framework used for modeling data and metadata must be specified. By explicitly stating the logical framework, such as description logics using OWL or first order logic using the Common Logic Interchange Framework, propositional interoperability is further promoted, enabling standardized data representations and query mechanisms. This sub-principle relates to Interoperability.
- **R1.4**: Metadata must specify the certainty level (i.e., level of confidence) of the semantic content and thus the information that their data contains, which is essential for proper reuse of data and for preventing phenomena such as citation distortion (63,64).

---

**Box 4 | The updated FAIR Guiding Principles and their association with layers of the EOSC Interoperability Framework and the different aspects of semantic interoperability discussed above. They include the original FAIR Guiding Principles (5) (in regular font) and proposed additions to them (in bold font).**

| To be Findable: | EOSC IF |
|---|---|
| F1. (meta)data are assigned a globally unique and persistent identifier | *technical* |
| F2. data are described with rich metadata (defined by R1 below) | *semantic* |
| F3. metadata clearly and explicitly include the identifier of the data it describes | *semantic* |
| F4. (meta)data are registered or indexed in a searchable resource | *technical* |
| F5. **vocabularies are used that support terminological interoperability** | ***semantic-terminological*** |
|   F5.1 **terms with the same meaning and reference/extension are ideally mapped across all relevant vocabularies through ontological and referential entity mappings** | ***semantic-terminological*** |
|   F5.2 **terms include multilingual labels and specify all relevant synonyms** | ***semantic-terminological*** |
| F6. **(meta)data schemata are used that support propositional interoperability** | ***semantic-propositional*** |
|   F6.1 **the same (meta)data schema is used for the same type of statement or collection of statement types, and the schema is referenced with its identifier in the statement's metadata** | ***semantic-propositional*** |
|   F6.2 **(meta)data schemata for the same type of statement are ideally aligned and mapped across all relevant schemata (i.e., schema crosswalks)** | ***semantic-propositional*** |
| F7. **(meta)data use a formalism to clearly distinguish between lexical, assertional, contingent, prototypical, and universal statements** | ***semantic-propositional*** |

| To be Accessible: | EOSC IF |
|---|---|
| A1. (meta)data are retrievable by their identifier using a standardized communications protocol | *technical* |
|   A1.1 the protocol is open, free, and universally implementable | *technical & legal* |
|   A1.2 the protocol allows for an authentication and authorization procedure, where necessary | *technical* |
|   A1.3 **the protocol is compliant with existing data protection regulations (e.g., [General Data Protection Regulation](#), GDPR)** | ***organisational & legal*** |
| A2. metadata are accessible, even when the data are no longer available | *technical & organisational* |

| To be Interoperable: | EOSC IF |
|---|---|
| I1. (meta)data use a formal, accessible, shared, and broadly applicable language for knowledge representation | *semantic & technical* |
| I2. (meta)data use vocabularies that follow FAIR principles | *semantic & technical* |
| I3. (meta)data include qualified references to other (meta)data | *semantic* |
| I4. **vocabularies used by (meta)data provide human-readable ontological definitions and, where applicable, human-readable recognition criteria (i.e., operational definitions) for their terms** | ***semantic-terminological*** |
| I5. **(meta)data specify the logical framework that has been used for their** | ***semantic-propositional*** |
| modeling (e.g., description logics using OWL or first order logic using Common Logic Interchange Framework)<br>I6.　　see F7.<br>I7.(1-2)　see F5.(1-2)<br>I8.(1-2)　see F6.(1-2) | *semantic-propositional*<br>*semantic-terminological*<br>*semantic-propositional* |
| **To be Reusable:**<br>R1.　　　(meta)data are richly described with a plurality of accurate and relevant attributes<br>　R1.1　　(meta)data are released with a clear and accessible data usage license<br>　R1.2　　(meta)data are associated with detailed provenance<br>　R1.3　　(meta)data meet domain-relevant community standards<br>　**R1.4　　metadata indicate the certainty level of the truthfulness of the semantic content of data** | EOSC IF<br>*semantic*<br><br>*legal*<br>*semantic*<br>*semantic & organisational*<br>*semantic* |

## FAIR Services

At their heart, the FAIR Guiding Principles provide recommendations for achieving rich machine-actionable data. Considering the challenges we discussed above regarding FAIRness and semantic interoperability, we must conclude that agreeing on a standard for FDOs that is only based on a set of minimum required metadata and on a specific format will not be sufficient for reaching this goal and for establishing the IFDS. Additionally, we must develop **FAIR Services** with which FDOs can be communicated and that support their FAIRness by indicating which operations can be conducted with a given FDO and which functions must be added to the Services for operations that are not yet supported for this type of FDO. We believe that such FAIR Services must comprise the following components:

1. A **Terminology Service** for supporting terminological interoperability. The service must comprise a repository and registry not only for controlled vocabularies, thesauri, taxonomies, and ontologies but also for entity mappings, with each mapping being documented as a FDO that can be referenced via its GUPRI and curated independent of any terminology. The entity mappings must distinguish ontological from referential mappings and other types of mapping relations (see Table 1) and should follow standards with detailed metadata (e.g., SSSOM, see Box 3). The service must also include a look-up service for terms and for entity mappings. Ideally, for each term, the service also provides information on the schemata in which it is used.

2. A **Schema Service** for supporting propositional interoperability. It must comprise a repository and registry not only for all kinds of data schemata (covering graph-based, tabular, and other kinds of data structures), organized based on the statement type(s) that the schema models, but also for schema crosswalks. Each schema and each crosswalk must be documented as a FDO that can be referenced via its GUPRI and curated independent of any other schema or crosswalk. Both schemata and crosswalks should follow standards for their documentation comparable to the SSSOM standard. Analog to the Terminology Service, the Schema Service must also include a look-up service for schemata and schema crosswalks. Since schemata for the same type of statement cannot only differ in the way they relate the different slots (i.e., syntactic positions) but also in the choice of vocabularies/ontologies for specifying constraints on slots (see Fig. 4), the Schema Service must utilize entity mappings and thus interact with the Terminology Service to be able to provide operational schema crosswalks.

3. An **Operations Service** for providing operations on FDOs. It must comprise a repository and registry for all kinds of functions (i.e., executable code) for operations that can be conducted on data, such as unit conversion. Each function must be documented as a FDO that can be referenced via its GUPRI and that can be associated with those schemata and vocabularies that it can be applied to by referencing their corresponding FDOs. Ideally, each schema, depending on its underlying data structure, has at least one example SPARQL, Cypher, or SQL query associated that is registered in this Operations Service.

If, in addition to such FAIR Services, the **FDOs of data statements or semantically meaningful collections of statements** would cover the following metadata, we would have a FAIR ecosystem at our disposal that could achieve a high degree of FAIRness for all kinds of data:

1. the GUPRI of the schema that has been applied to structure the content of the FDO, with the schema itself being described, registered, and made available as a schema FDO in a schema repository;
2. distinction between the creator of the FDO and the author(s) of its contents;
3. specification of the type of statement(s) the FDO contains (i.e., lexical, assertional, contingent, prototypical, universal);
4. specification of the formal logical framework that has been applied in the FDO, if any, to inform whether one can reason over its contents and which logical framework must be used for it;
5. a human-readable representation of the content to meet requirements of cognitive interoperability (21);
6. specification of the degree of certainty of the contents of the FDO.

Agreeing on a minimum metadata standard for FDOs would increase the **accessibility** of data in the IFDS. However, for the IFDS to be truly FAIR, the FAIR Services would also have to provide the entity mappings and schema crosswalks operationally, supporting not only the applicability of operations across different schemata and terminologies (**interoperability**) and facilitating data integration and thus data **reuse** in the IFDS, but also increasing the general **findability** of data across different platforms and repositories in the IFDS, independent of the particular terminologies and data schemata used. As a result, **FDOs that are supported by FAIR Services as described above would represent units of interoperability in the IFDS**.

---

**Box 5 | Existing Work on FAIR Services**

A number of **terminology services** exist, some of which are registries, such as Linked Open Vocabularies (LOV), the BioRegistry or Archivo, that mostly provide metadata on the terminology level, while others are look-up services, such as EMBL-EBI Ontology Look-Up Servcie (OLS) and the TIB Terminology Service that is based on it, or NCBO BioPortal and related OntoPortal based domain repositories, that also provide metadata on the term level and additional features for browsing, displaying term mappings (e.g., generated by the LOOM algorithm), or managing the indexed terminologies. There is also Skosmos (65), an open source web-based browser and publishing tool specialized for Simple Knowledge Organisation System (SKOS) vocabularies. The Basic Register of Thesauri, Ontologies & Classifications (BARTOC) is a database of Knowledge Organization Systems and related registries which collected and lists over 100 terminology registries. Many of the latter are often focused on indexing terminologies of specific research areas only, such as Biology, Chemistry, or Medicine. Consequently, there is the need to use terminology metadata schema standards, such as the Metadata for Ontology Description and Publication Ontology (MOD), for the interoperability between all of these terminology services with regard to synchronizing terminology and mapping metadata for interdisciplinary contexts and use cases.

> **EMBL-EBI Ontology Xref Service (OxO)** (66) is a service designed to find mappings or cross-references between terms from various ontologies, vocabularies, and coding standards. OxO imports these mappings from multiple sources, including OLS and a subset of mappings provided by the UMLS (Unified Medical Language System). The service is still under development, and users are encouraged to provide feedback. OxO allows users to search for mappings using specific identifiers, or to view all mappings between different data sources.
>
> The [Metadata Schema and Crosswalk Registry (MSCR)](#) (67) and the **Data Type Registry (DTR)** of the [FAIRCORE4EOSC](#) project are registries for managing and sharing metadata schemata and crosswalks respectively data types within the EOSC ecosystem. It allows users to search, browse, and download metadata schemata and crosswalks or data types via a GUI or an API. Registered users and communities can create, register, and version metadata schemata and crosswalks or data types with GUPRIs.
>
> The Data Repository Service (DRS) API (68,69) provides a generic interface to data repositories so data consumers, including workflow systems, can access data objects in a single, standard way regardless of where they are stored and how they are managed. The primary functionality of DRS is to map a logical ID to a means for physically retrieving the data represented by the ID.
>
> [FAIRsharing.org](#) (70) is a community-driven platform, with a diverse set of stakeholders from academia, industry, funding agencies, standards organizations, and infrastructure providers. It aims to increase guidance for consumers of standards, databases, repositories, and data policies, as well as improve producer satisfaction in terms of resource visibility, reuse, adoption, and citation.
>
> [Mapping Commons](#) (57) is an idea developed and promoted by the Monarch Initiative (71) and the SSSOM Developer Community, which involves the creation and maintenance of domain-focused, community-curated mapping registries. A template system has been developed to support setting up and maintaining registries, with some support for managing the mapping life-cycle (in particular data model validation).

For the implementation of such FAIR services, existing and ongoing work (see Box 3 & Box 5) must be considered for reuse and build upon whenever possible. Open-source based collaboration between the associated, international stakeholders of FAIR initiatives and projects needs to be intensified, championed and properly acknowledged by funders. Although the IFDS is inherently decentral, we believe that to ease and secure FAIRness, institutions like libraries should function as trusted focal points within it, by bundling all three FAIR Service components and periodically harvesting (meta)data from other IFDS nodes to index, archive, and serve relevant FDOs as common good for humans and machines persistently.

# Conclusion

In an Internet of FAIR Data and Services (IFDS) that is based on a FAIR ecosystem as the one described above, any person, including researchers and scientists, could refer to any particular FAIR Digital Object (FDO) and make statements about it. FDOs, when used to organize semantically meaningful subsets of given datasets (see *semantic units*, (61)), can cover information at different levels of granularity, with individual statements representing the finest and collections of statements representing coarser levels of granularity. Applying FDOs in this nested way could have far-reaching consequences and fundamentally change the way we publish and communicate research in the future. Authorship of scientific contributions could be specified at the level of individual statements, which would contribute to a more fair reflection of the actual work that has gone into a scientific contribution by each person involved, as opposed to the list of authors of a classical scientific paper.

Also, smaller contributions that may not be sufficient for publication in a paper could be published as (a set of) FDO(s). Instead of citing a paper as a whole, in the IFDS, researchers could cite the FDO they want to refer to and make a statement about it, which, in turn, would be documented as an FDO and which would be a citation. Since the citing FDO would specify in its provenance metadata who is citing which other FDO, and since its content would specify the nature of that citation, one could classify the citation (e.g., as supporting or contradicting the cited FDO), resulting in qualified references that would specifically target the content to which they refer, rather than entire papers.

If this approach were to become common practice in academia, it would likely change the way researchers build their careers, as a researcher's qualification could be quantitatively assessed in terms of their FDOs and the number of citations made to them by other researchers. The citations could be weighted and evaluated differentially (i.e., supporting versus contradicting citations) and could replace the number of paper citations weighted by the journal's impact factor. Researchers could build a successful academic career without the need to publish in highly ranked journals if their statement FDOs were highly referenced by other researchers. As a result, high-impact journals would lose some of their appeal to researchers, and academia could finally emancipate itself from the grip of the publishing industry, and the IFDS would become a general communication platform for all researchers.

# Acknowledgements

We thank Björn Quast, Peter Grobe, István Míko, Kheir Eddine, Daniel Mietchen, Felix Engel, and Christian Hauschke for discussing some of the presented ideas. We are solely responsible for all the arguments and statements in this paper. Lars Vogt was financially supported by the ERC H2020 Project 'ScienceGraph' (819536) grant and by the project DIGIT RUBBER funded by the German Federal Ministry of Education and Research (BMBF), Grant no. 13XP5126B (TIB). Philip Strömert was financially supported by the [Deutsche Forschungsgemeinschaft (DFG, German Research Foundation)](#) under the [National Research Data Infrastructure – NFDI4Chem](#) (Projektnr.: 441958208).